\documentclass[12pt]{article}

\usepackage{graphicx}
\newcommand {\bi} {\bibitem}
\newcommand {\be} {\begin{equation}}
\newcommand {\ee} {\end{equation}}
\newcommand {\bea} {\begin{eqnarray} }
\newcommand {\eea} {\nonumber \end{eqnarray}}
\newcommand {\eps} {\epsilon}
\newcommand {\si} {\sigma}

\newcommand {\ba} {\overline}
\newcommand {\lan} {\langle}
\newcommand {\ran} {\rangle}

\newcommand {\bc} {\begin{center}}
\newcommand {\ec} {\end{center}}
\newcommand {\bd}{\begin{displaymath}}
\newcommand {\ed}{\end{displaymath}}

\def \form#1 {eq. (\ref{#1}) }
\def \parziale#1#2  {{\partial {#1} \over \partial {#2}}}
\def \bi#1 {\typeout{#1} \item}

\begin{document}

\title{Spin Glass Theory: numerical and experimental results in three-dimensional systems}

\author{Giorgio Parisi \\
Dipartimento di Fisica, Sezione INFN, SMC and UdRm1 of INFM,\\
Universit\`a di Roma ``La Sapienza'',\\
Piazzale Aldo Moro 2,
I-00185 Rome (Italy)\\
Giorgio.Parisi@Roma1.Infn.It}
\maketitle
\begin{abstract}
	 Here I will review the theoretical results that have been obtained for spin glasses.  I
	 will concentrate my attention on the predictions of the mean field approach in three dimensional
	 systems and on its numerical and experimental verifications.  
\end{abstract}

\section{Introduction}

{Spin glasses have been intensively studied in the last thirty years}. They are the simplest example of glassy
systems. There is a highly non-trivial mean field approximation, that can be used to derive some of the main
properties of glassy systems\cite{ANDERSON,MPV,PARISIB,CINQUE}.
{The study of spin glasses opens an important window for studying off-equilibrium behavior.} Aging
\cite{Bou}
and the  violations of the equilibrium fluctuation dissipation relations emerge in a natural way.
\cite{Cuku1,FM,FMPP}.  
Many of the ideas developed in this context  have a wide domain of applications.

{The simplest spin glass Hamiltonian} is of the form:

\be H=\sum_{i,k=1,N}J_{i,k} \si_{i}\si_{k}\, , 
\ee 
where the $J$'s are {\it quenched}  random variables located on the links  connecting
two points of the lattice,
the $\si$'s are {Ising} variables (i.e. $\pm 1$).  
The total number of points is denoted with $N$ and
it goes to infinity in the thermodynamic limit.

The most studied models are
\begin{itemize}
\item
{The SK model:}
\cite{SK}
All $J$'s are random and different from zero, with a Gaussian or a bimodal
distribution with variance $N^{-1/2}$.  The coordination number is $N-1$ and it goes to infinity with
$N$.  {In this case a \emph{simple} mean field theory is valid} in the infinite $N$ limit.
\cite{MPV,PARISIB}.
\item
{The Bethe lattice model:}
 \cite{VianaBray,MP1Be,FL}: 
The spins live on a random lattice and only $Nz/2$ $J$'s are
different from zero: they have variance $z^{-1/2}$.  The  coordination number is 
$z$.  {A modified mean field theory is valid.}
\item
{The Edwards Anderson model:}
 \cite{EA}: 
The spins belong to a finite dimensional lattice of dimensions
$D$: Only nearest neighbor interactions are different from zero and their variance is $D^{-1/2}$.
In this case {finite corrections to mean field theory are present, that are certainly very large in
one or two dimensions.}
\end{itemize}

{The SK model is the limit of the EA model when the dimension goes to infinity and it is a good starting
point} for studying also the finite dimensional case with short range interaction, that is the most realistic
and difficult case.

\section{Equilibrium properties in the mean field approximation}

Let us describe some of the equilibrium properties that can be  computed in the mean field
approximation.
At low temperature at equilibrium a large, but finite, spin glass system {remains for an exponentially large
time in a  small region of phase space}, but it may jump occasionally in a relatively short time to an
other region of phase space.
It is like the theory of {\it punctuated equilibria}: long periods of
\emph{stasis}, punctuated by fast changes.
We call  each of these regions  of phase space an {\it equilibrium state} or {\it quasi-equilibrium state}.
We label with $\alpha$ each state and we
can define a local magnetization:
{
\be
m(i)_{\alpha}=\lan\si_{i}\ran_{\alpha}\, .
\ee
}

{When the mean field theory is valid}, this magnetization satisfies the mean field equations that in a first
approximation can be written
neglecting the Bethe-TAP reaction-cavity term \cite{TAP}, can be written (in the SK limit) as
{
\be
m(i)=\mbox{th}(\beta h(i)) \, \ \ \
 h(i)=\sum_{k}J_{i,k}m(k)\, . 
\ee
}
It is non-trivial to compute how many solutions this equation has and which
are the properties of the solutions.  
{The number of solutions 
is exponential large \cite{BMD}} and the precise
computation of its value is a complex task 
 \cite{R1,R2,R3,R45,R5,R55,R6}.
 
The $\alpha$-th
solution is present in the dynamics at equilibrium for a time proportional to 
\be 
w_{\alpha}\propto \exp
(-\beta f_{\alpha})\, , 
\ee 
where {$f_{\alpha}$ is the total free energy} that can be computed from the
magnetizations using an explicit formula.

In general one finds that
\be
P(f) \approx \exp(N\Sigma(f/N))\, .
\ee
where $\Sigma (F)$ is  a non trivial function called the complexity.
 \cite{MONA}. 
Near the ground state the previous formula simplifies to
\be
P(f)\propto \exp( y (f-f_{0}))\ , 
\ee
where
$f_{0}=N F_{0}$,  the complexity satisfied the condition  
$\Sigma(F_{0})=0 $ and
${d\Sigma(F)\over dF} {|}_{F=F_{0}} \equiv y<\beta$.

Although the number of states exponentially increases with the free energy, this increase is slower that
$\exp(\beta f)$, and {statistical sums are dominated by the lowest free energy states}: a small
number of states carries most of the statistical weight.

The states are macroscopically different: {  it is 
convenient to  define the macroscopic distance $d$ and the overlap
$q$:}
\bea
d(\alpha,\gamma)^2={\sum_{i}\left( m(i)_{\alpha}-m(i)_{\gamma}\right) ^2\over N}\ , \\
q(\alpha,\gamma)={\sum_{i}m(i)_{\alpha}m(i)_{\gamma}\over N} \ . \label {DIS}
\eea
{The states are equivalent: intensive observables (that depends only on a single state)
have the same value in all the states}, e.g. the self overlap does not depend on the state:
\be
q(\alpha,\alpha)=q_{EA}\ \ \ \forall \alpha \ .
\ee

Obviously distance and overlap are related; 

\be
d(\alpha,\gamma)^2=2 q_{EA}-2 q(\alpha,\gamma).\ee
For historical reasons this picture is called {\emph{replica symmetry breaking}}.

 \begin{figure}
  \includegraphics[width=0.33\textwidth,angle=270]{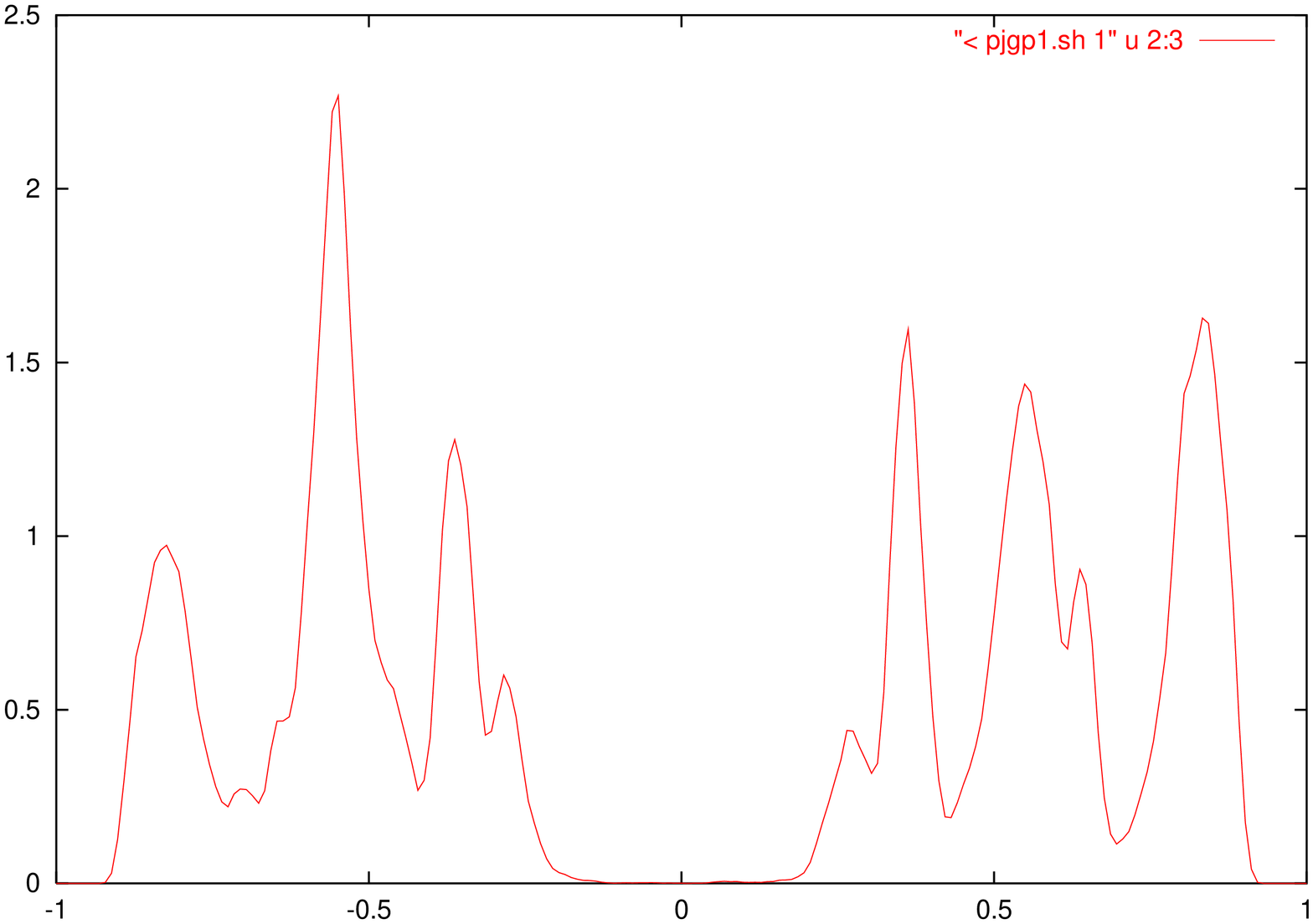}
  \includegraphics[width=0.33\textwidth,angle=270]{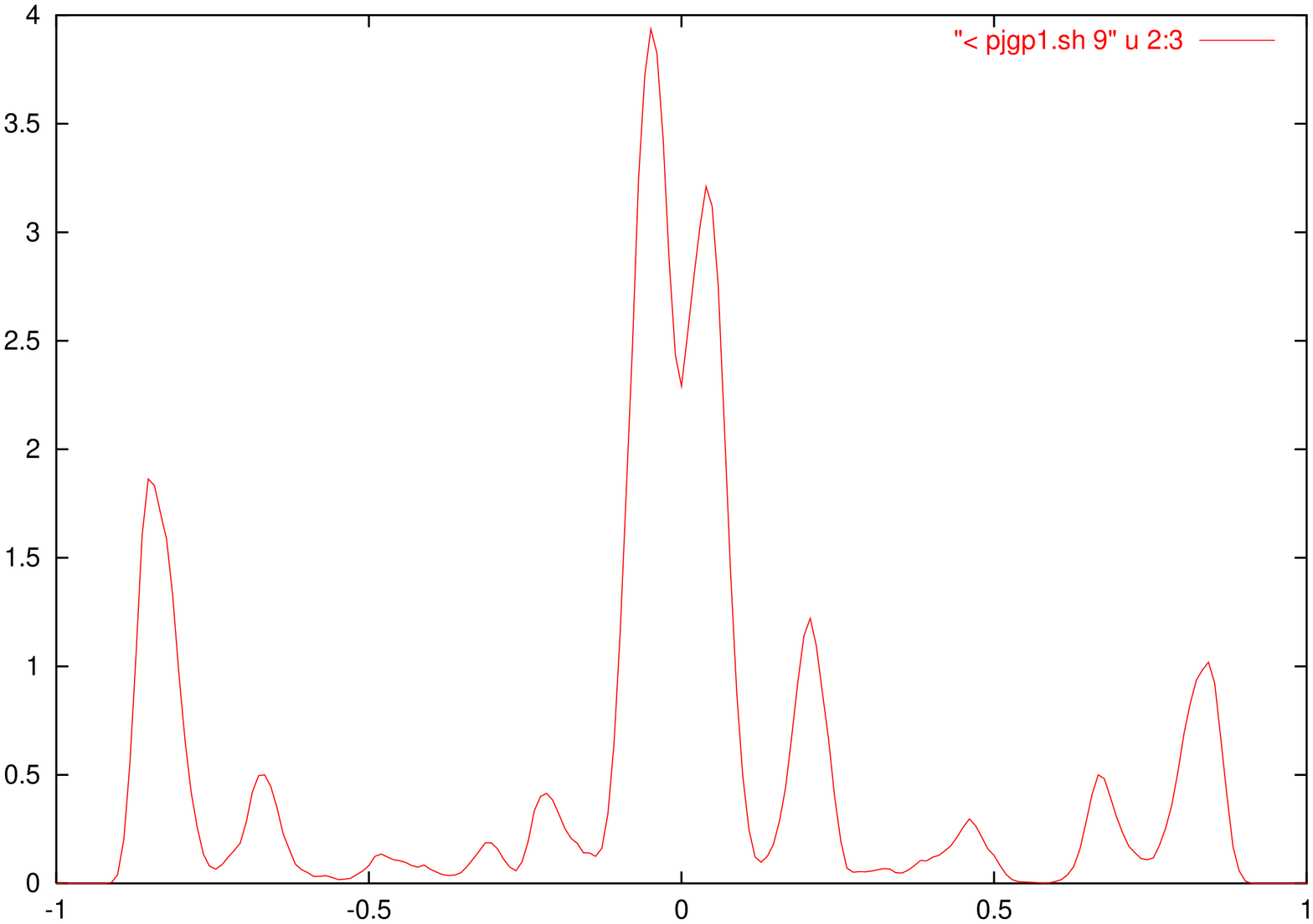}
  \caption{
  The function $P_{J}(q)$ for two  
  samples (i.e two choices of $J$) for $D=3,\  L=16$ ($16^{3}$ spins).
   from \cite{NUM1}.}
 \label{A}
 \end{figure}

For each given system it is convenient to introduce  the function $P_{J}(q)$, i.e. {the
probability distribution of the overlap among two equilibrium configurations}, see fig. (\ref{A}).
Using the metaphor of having many equilibrium states, we can  write for large systems
\be
P_{J}(q)= \sum_{\alpha,\gamma}w_{\alpha}w_{\gamma}\delta(q_{\alpha,\gamma}-q) \ .
\ee

\begin{figure}
  \includegraphics[width=.50\textwidth]{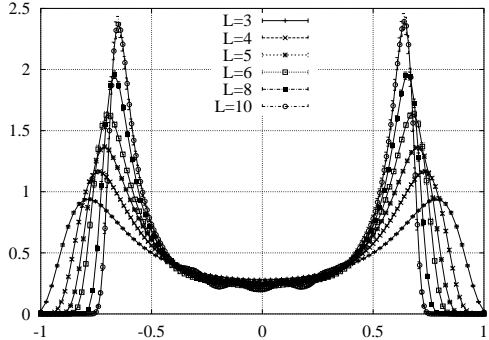}
   \caption{
  The function $P(q)=\ba{P_{J}(q)}$ after average over many samples (L=3\ldots 10).
   from \cite{NUM2}. }
 \label{AVE}
 \end{figure}

We define  
{
\be
 P(q) \equiv \overline {P_{J}(q)} \ee
}
The average is done over the different
choices of the couplings $J$, see fig.  (\ref{AVE}).  

This average is needed because the theory
predicts (and numerical simulations also in three dimensions do confirm) that {the function
$P_{J}(q)$ changes dramatically from system to system.}

In the mean field approximation the function $P(q)$ (and its fluctuations from system to system) {can
be computed analytically:} at zero magnetic field $P(q)$ has two delta
functions at $\pm q_{EA}$, with a flat part in between.

 {Very interesting phenomena happen when we add a very small magnetic field.}
The order of the states in free energy is scrambled: their free energies differ of a factor $O(1)$ and the
perturbation is of order $N$. Different results are obtained if we use different experimental protocols:
\begin{itemize}
    \item
    {If we add the field at low temperature}, the system remains for a very large time in the same state, only
asymptotically it  jumps to one of the lower equilibrium states of the new Hamiltonian.
\item
{If we cool the system from high temperature in a field}, we likely go directly to one of the 
good lowest free energy states.
\end{itemize}

{Correspondingly there are two susceptibilities that can be measured also experimentally:}
\begin{itemize}
    \item
    {The so called linear response susceptibility $\chi_{LR}$}, i.e. the response within a state, that is
observable when we change the magnetic field at fixed temperature and we do not wait too much.  
This
susceptibility is related to the fluctuations of the magnetization inside a given state.
\item
{The true equilibrium susceptibility, $\chi_{eq}$}, that is related to the fluctuation of the
magnetization when we consider also the contributions that arise from the fact that the total
magnetization is slightly different (of a quantity proportional to $\sqrt{N}$) in different states.
This susceptibility  is very near to { $\chi_{FC} $, the field cooled susceptibility, where one
cools the system in presence of a field.}
\end{itemize}

\begin{figure}
	 \includegraphics[width=.49\columnwidth]{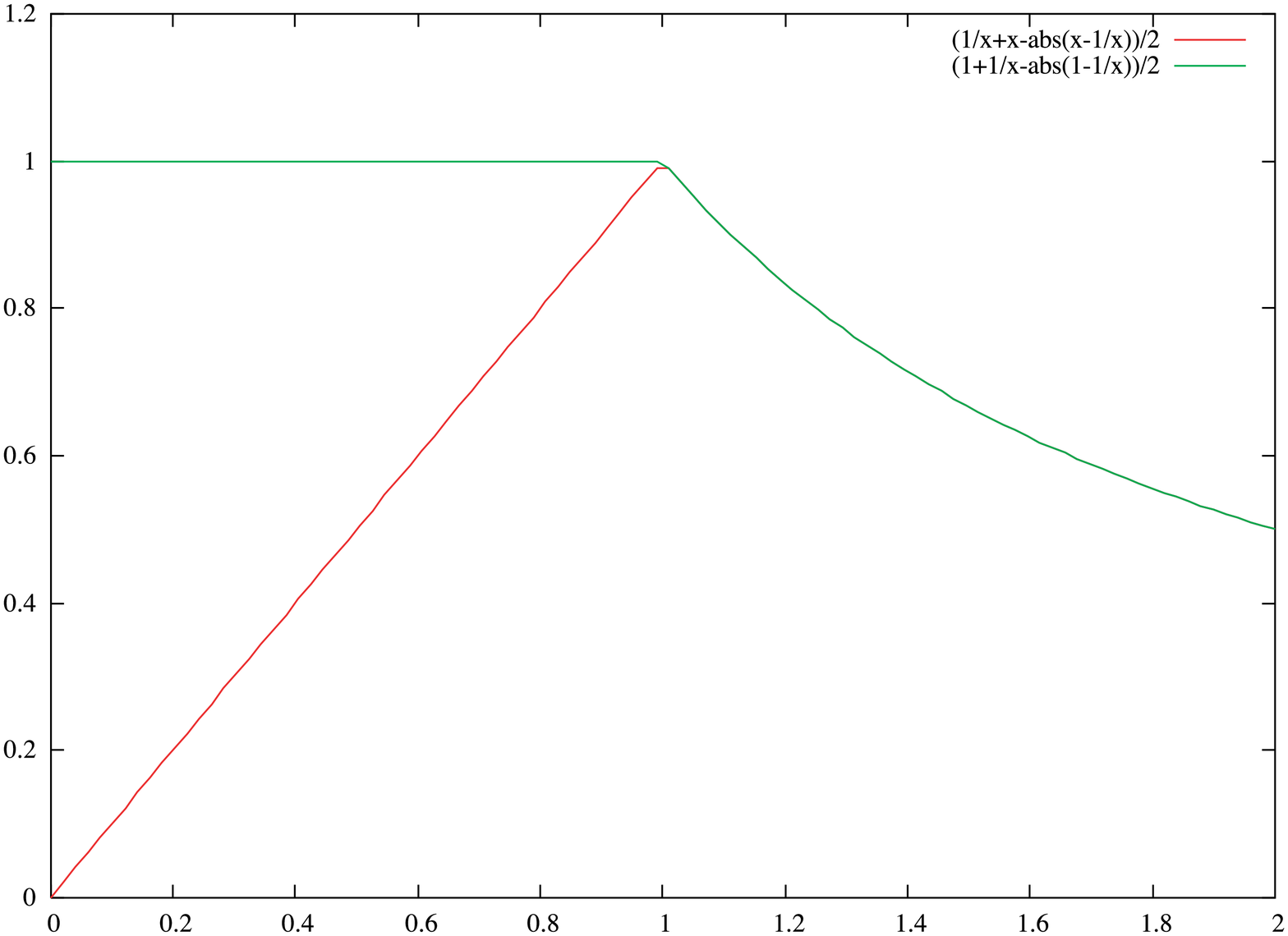}
	 \includegraphics[width=.49\columnwidth]{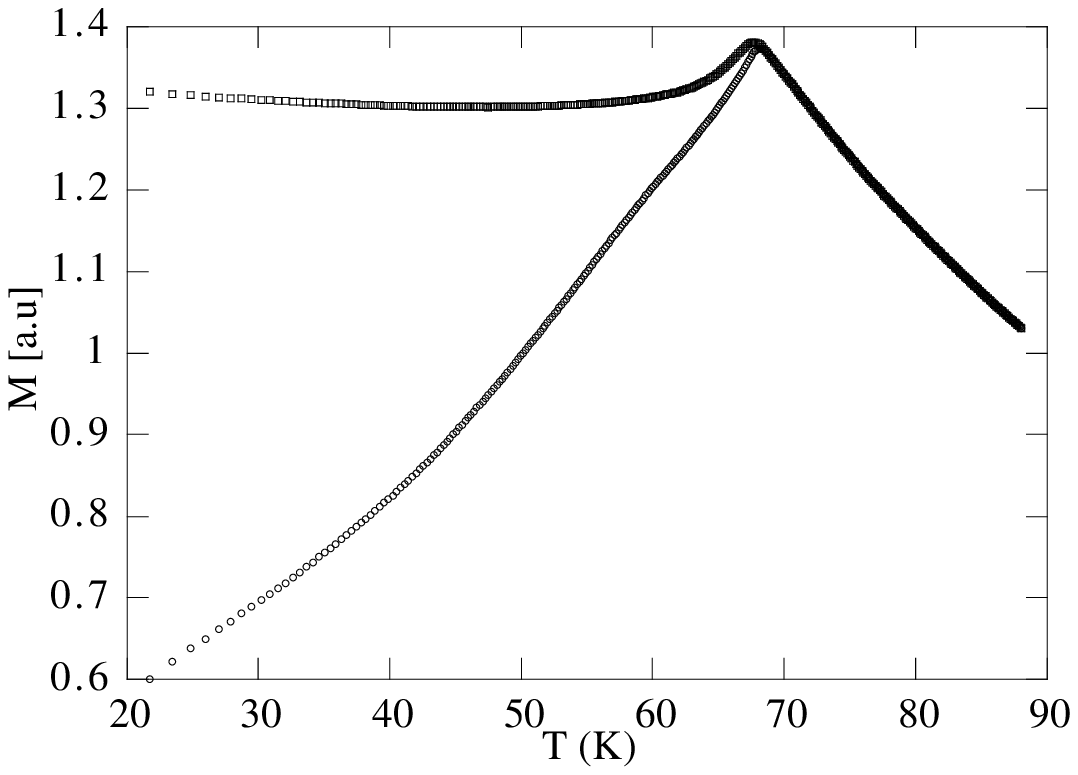}
 	 \caption{Right panel, the analytic results in the mean field approximation.}
	 \cite{MPV};
	 {left panel, the experimental results for a metallic spin glass.	 }
 	 \cite{EXP1}

 	 \label{TWOS}
 	 \end{figure}

{The difference of the two susceptibilities is the hallmark of replica symmetry breaking.}
 In fig. (\ref{TWOS}) we show both the analytic results for the SK model 
 \cite{MPV} 
and  the experimental data on
metallic spin glasses 
 \cite{EXP1}. 
The similarities among the two panels are striking.

This phenomenon is quite different from hysteresis.
Hysteresis is due to defects that are localized in space and produce a finite barrier in
	 free energy.  {The mean life of the metastable states is finite} and it  is roughly $\exp (\beta
	 \Delta F)$ where $\Delta F$ is a number of order 1 in natural units.
On the contrary in the mean field theory of spin glasses the system must cross barriers that correspond to
	 rearrangements of arbitrary large regions of the system. {The largest value of the barriers diverge 
	 in the thermodynamic limit.}
{In hysteresis, if we wait enough time the two susceptibilities  coincide}, while
	 {they remain always different in this new framework} if the applied magnetic field is
	 small enough (non-linear susceptibilities are divergent).

\section{Slightly off-equilibrium behavior}

The theoretical study of non-equilibrium systems is not easy.  However some detailed results can
be obtained if the system is slightly off-equilibrium.  
A neat theory may be formulated when {the time scale related to non-equilibrium phenomena is much longer than
the microscopic time scales }(e.g seconds versus picoseconds).  The simplest way to put a system out of
equilibrium it to perturb it by changing the external parameters (temperature, magnetic field) and to produce a
transient behavior.

\subsection{Aging}
{The mostly studied case is aging. }
The system is cooled from a high temperature to a low
temperature at time 0.  
Aging implies that the response at a {time scale (i.e. inverse frequency) that is of
the order of the age of the system} is notably different from the equilibrium one. 
One time quantities (i.e. internal energy) converge much faster to equilibrium (may be with a power law
behavior).

In the case of {\sl naive aging} at large time, at very low frequency ($\omega$), in the region where
\be
\omega t<<1\ ,
\ee
the dependence of the response scales as
\be
\chi(t,\omega)=\hat \chi(\omega t) \ .
\ee
Part of the susceptibility receive a contribution from processes that takes a time $t$ to happen,
where $t$ is the age of the system.

\subsection{A possible origin of aging}

When we cool the system below the critical temperature, the system has the tendency to order itself,
i.e. to go to one of the equilibrium states (al least two in absence of a magnetic field).  

This process  happen locally (there is no direct long
range exchange of information) and the degrees of freedom arrange themselves in  some
configurations that locally minimize the free energy.  In this process {we have the formation of
domains where the free energy is well minimized} separated by walls with higher free energy.
Therefore at finite time we have the formation of a mosaic state,
characterized by a
dynamical length $\xi(t)$.  The function $\xi(t)$ is an increasing function of time that {eventually
goes to infinity.}

While in the case of spinoidal decomposition there are only two equilibrium states, and the mosaic
has only two colors, {when there are many locally different equilibrium states, as it
should be in spin glasses, each cell of the mosaic is likely to belong to a different ground state}
and the picture is much more complex.

In the case of the spinoidal decomposition the function $\xi(t)$ increases relatively fast (e.g. as
$t^{1/3}$).  
In spin glasses, the increase of the function $\xi(t)$ is rather slow.
There are indications that also in the most favorable experimental situations { $\xi(t)$ arrives to 100 (in
microscopic units)} \cite{NOI,ORBACH}: finite size effects in the dynamics disappear from system with more than $10^6$ spins.
{The precise form of
the increase of the function $\xi(t)$ is not very important.}
There are suggestions that it increase as 
\be
t^\alpha \ ,
\ee
{with $\alpha$ of the order 0.13} in some experiments done at $T/ T_{c} \approx .75$.

The excess of energy \cite{NUM3} is proportional to a high negative power of $\xi(t)$ (e.g. as
$\xi(t)^{-4}$).  During aging energy relaxation is very small: it has never been observed
experimentally, but only seen in simulations (there is a better control at short time).  In this situation {the
system moves microscopically much more than at equilibrium}, because when $\xi$ increases different
domains are rearranged and this produces an excess of thermal fluctuations.

In
the same way {the systems may choose among different possibilities} when the domains change and this
may lead to an additional response to external perturbations that may influence these choices.
During aging {the relations between fluctuations and response are modified. }

\subsection{Generalized fluctuation dissipation relations}
The  fluctuation dissipation
theorem, that is at the basis of the thermodynamics and  is a consequence of the so called
zeroth law of the thermodynamics, is no more valid: {a new definition of \textit{temperature} is needed.}

{Let us show how these ideas are implemented for the aging of spin glasses.}  Our aim is to define a
correlation function and a response function in a consistent way such that the new off-equilibrium
fluctuation dissipation relations can be found.
The correlation function of total magnetization is defined as
{
\be
C(t,t_w) \equiv \lan m (t_w) m( t_w+t)\ran\ .
\ee
}
In spin glasses at zero external magnetic field  the off-diagonal terms averages
to zero and the only surviving term is 
\be
C(t,t_w) =\frac1N \sum_{i=1}^N \lan \sigma_{i} (t_w) \sigma_{i}( t_w+t)\ran =  q(t_{w},t_{w}+t) \ ,
\ee
i.e.  the overlap $ q(t_{w},t_{w}+t)$ between a configuration at time $t_{w}$ and one at 
time $t_{w}+t$. 

 \begin{figure}
\includegraphics[width=0.75\textwidth]{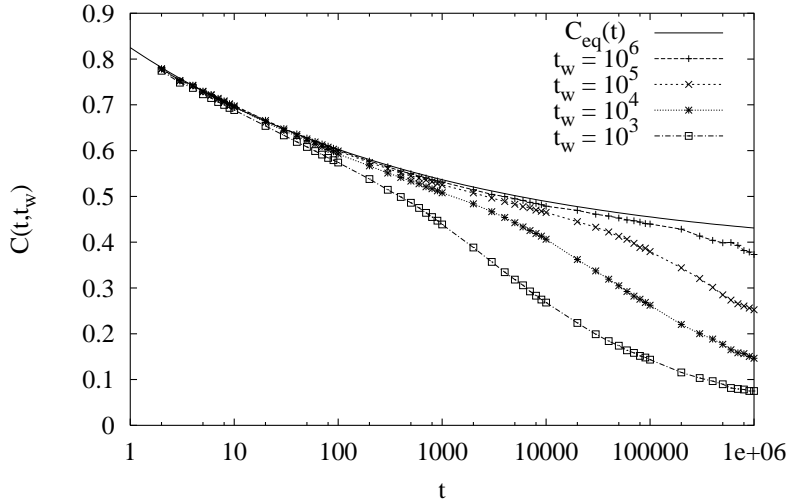}
 \caption{
The correlation function for spin glasses  as function of time $t$ at 
different $t_{w}$ from simulations .
 \cite{NUM3}).
}
 \label{CORR}
 \end{figure}

The relaxation function $S(t,t_w)$ is  just given by
\be
S(t,t_w)\equiv R(t,t_w) =\beta ^{-1}\lim_{\delta h \to 0 } {\delta \lan m(t+t_{w}) \ran \over \delta h}\ ,
\ee
where $\delta m$ is the variation of the magnetization when we add  a magnetic 
field $\delta h$ starting from time $t_{w}$ and
$S(t,t_{w)}$ and $\chi(t_{w},\omega)$ are related by Fourier transform.

Naive scaling implies
\be
C(t,t_{w)}=\hat C(t/t_{w}), \ \ \ \ \ \ R(t,t_{w)}=\hat R(t/t_{w})\, 
\ee
The dependence on $t$ and $t_{w}$ of the previously defined functions is rather complex and {cannot be computed
from general principles. }

It is convenient to examine directly the relation between $S$ and $C$, by eliminating the
time.  At this end {we plot parametrically $S(t,t_{w})$ versus $C(t,t_{w})\equiv
q(t,t_{w})$ at fixed $t_{w}$.}
The theory predicts that such a  plot goes to a finite limit when $t_{w} \to \infty$ and we can extract from it
information on the phase structure of equilibrium configurations.
Using general arguments 
 \cite{Cuku1,FM,FMPP} 
one finds that when $t_{w}\to \infty$,
\be
{dS \over dC} = X(C) =\int_{0}^C dq P(q) \ . \label{FDR}
\ee
This  equation is very important because it connects quantities that can be measured in the dynamics (l.h.s)
to quantities that a defined at equilibrium (r.h.s.).
\subsection{Three possible forms of fluctuation dissipation relations}
 \begin{figure}
\includegraphics[width=0.50\textwidth]{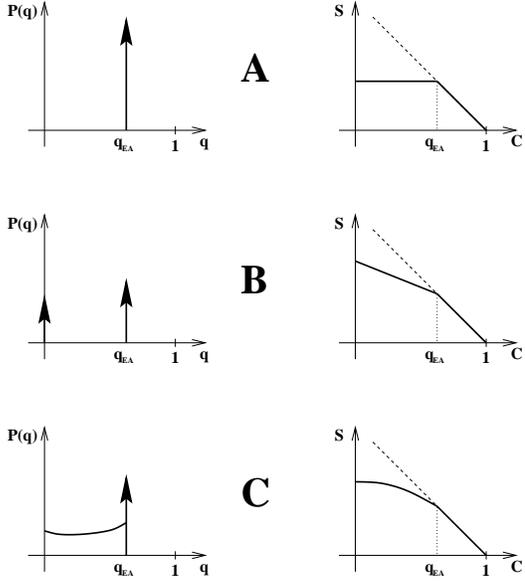}

 \caption{
Three different form of the function $P(q)$ (left) and the related function $S(q)$ (right).
 (taken from \cite{FIGURA}). 
}
 \label{3REP}
 \end{figure}

The behavior of the system at equilibrium and the modification of the fluctuation dissipation
theorem off-equilibrium are strongly related.  
These relations are summarized in fig. (\ref{3REP}).
In all the right panels the time decrease from right to left.  
{At short times, the relaxation  is a straight line
(with slope -1), according to the fluctuation dissipation theorem.  }
The interesting part is the one
at left, where at large times, in the aging regime, the curve deviates from the previous straight
line.  {The value of the relaxation at the point where the equilibrium regime ends is the linear
response susceptibility $\chi_{LR}$, while the value of the relaxation on the left-most point is the
equilibrium susceptibility $\chi_{eq}$.  }
\begin{itemize}
\item A:  There is an excess of noise with respect to equilibrium.  The presence
of noise without a corresponding response implies that {the effective temperature $\beta X(C)$ is infinite}.
\item B: A new and non trivial phenomenon, that should be present
in structural fragile glasses and in some kind
of spin glasses. {The system has two temperatures, and both are finite.}
 \cite{CKP,FV,LOCO3}. 
 Often the higher temperature 
 is near to the critical temperature.
\item C: A more complex phenomenon that  is present in the mean field theory of some spin glasses.
 (e.g. in the original SK model).
It correspond to { the presence of a continuous range of temperatures in the aging region.}
\end{itemize}

 \begin{figure}
    \includegraphics[width=.55\textwidth]{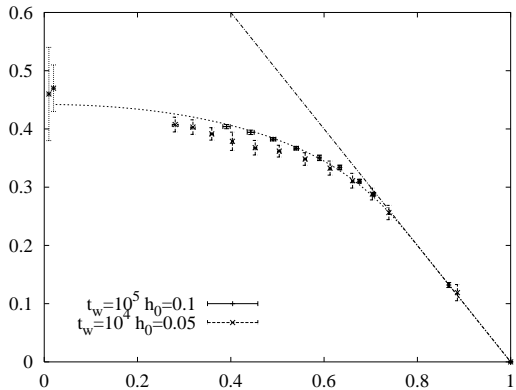}
 \caption{
Relaxation function versus correlation in the Edwards-Anderson (EA) model in $D=3$
	  $T=0.7\simeq \frac34 T_c$ and theoretical prediction (Ising case) \cite{NUM3}, obtained using equation
	  (\ref{FDR}).}
 \label {S}
 \end{figure}

The presence of anomalies in the off-equilibrium regime in the plot of the response versus correlations
(i.e. cases B and C) marks in a clear way the difference from the old picture (hysteresis).  The
experimental fact that in real spin glasses $\chi_{LR}\ne \chi_{eq}$ excludes case A.

In the case of Ising spin glasses with two body interaction in the mean field approximation we stay
in case C. 
What happens in three dimensions is not clear.  {Numerical simulations
 \cite{NUM1,NUM2,NUM3,CINQUE} 
on Ising model indicates that we stay in case C}, {while  the
experiments \cite{OCIO} 
 see a clear effect of deviations from the case A that may  indicate more case B. }
 
{From the theoretical
viewpoint there are no firm commitments}: although in infinite dimensions with a two-spin interaction
we are in case C, the corrections due to the interaction among the fluctuations could bring the
system in three dimensions in case B.

\subsection{The physical meaning of the fluctuation dissipation relations}
 {I will sketch a proof of the standard equilibrium fluctuation dissipation theorem using very simple and
very general arguments.}

{The basic tools are equipartition law for harmonic oscillators and the zeroth law of thermodynamics; it
states that:}
\begin{itemize}
	 \item Two bodies at thermal contact will eventually  have the same temperature.
	 \item If $A$ has the same temperature of  $B$ and $B$ has the same temperature of $C$,
	 $A$  has the same temperature of  $C$.
	 \item The heat goes from the body at higher temperature to the one at low temperature.
\end{itemize}
The equipartition law for harmonic oscillators states that the internal energy of an  one dimensional harmonic oscillator is
$ kT$.

{Let us consider an harmonic thermometer. }
We have a spring coupled to a a quantity $M(t)$ of the system whose temperature we want to measure.
The Hamiltonian of the thermometer is 
\be
H=\frac12(p^{2}+\omega^{2}x^{2})+\eps \; x M(t) \ ,
\ee
and $\eps$  is very small (and the thermalization time diverges as $\eps^{-2}$).
{We want to compute the asymptotic value of the energy of the thermometer. }

In order to be more precise, we can repeat the
experiment many times and do the average over the different experiments. {At $\eps=0$ we can define the correlation functions $C$}:

\be \lan M(t) M(t')\ran  = C(t,t') = C(t-t')\ .
\ee
The last equality is true if the system is in a stationary state.

At the first order in $\eps$ we have
\be
\lan M(t) \ran =\eps \int_{0}^{t}dt' x(t') r(t',t)\ ,
\ee
{where $r$ is the response function and $r(t,t') = r(t-t')$ in a stationary state.}

A long, but conceptually simple, computation tell us that the energy of the thermometer is $\omega$ independent 
and it is the correct one, if and only if we have the fluctuation dissipation relation:
{
\be
r(t)=-\beta{dC(t)\over dt} \ .
\ee
}

{If we introduce the relaxation function
\be
R(t)=\int_{0}^t r(t')dt' \ ,
\ee}
the fluctuation dissipation theorem becomes
\be
{dR(t)\over dt}=-\beta{dC(t)\over dt} \ .
\ee

It may be convenient to eliminate the time and consider $R$ as a function of $C$, e.g. to plot parametrically
$R(t)$
as function of $C(t)$. In this way the classical fluctuation dissipation theorem has a very compact form:
{
\be
{dR\over dC}=-\beta \ .
\ee}

\begin{figure}
    \includegraphics[width=.40\textwidth]{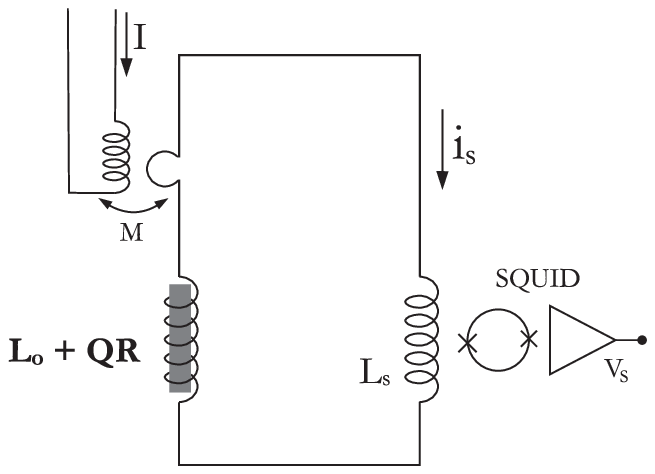} \includegraphics[width=.42\textwidth]{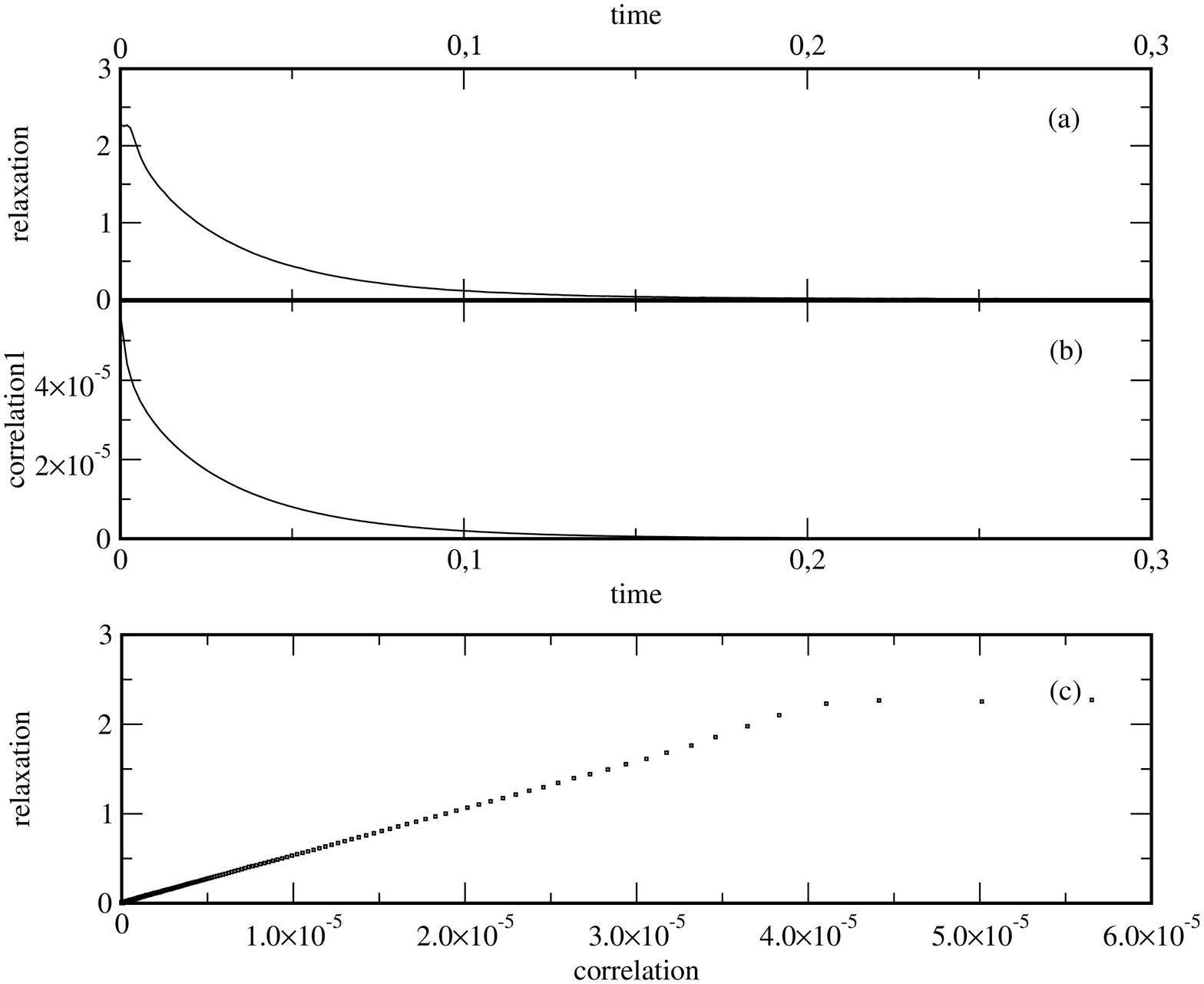}
    
\caption{
On the left: The schematic of a circuits used o measure the magnetic response and the correlations.
On the right: Measured relaxation (a) and autocorrelation (b) function for
the copper sample at 4.2K. (c) FD-plot, relaxation versus response,
the observation time $t=$ being used as parameter: the observed
linear behavior allows calibration of the system as thermometer \cite{OCIO}. The relaxation differs of by a sign and by an
additive
constant from the one used in the text.
}
 \label{3FLUBIS}
 \end{figure}

{Around 1993 \cite{Cuku1,FM} it was showed that in some slightly off-equilibrium
situations (e.g. during the ageing of the system) there is a generalized fluctuation dissipation relation.}
\be
{dR(t',t)\over dt}=-\beta X(C(t',t)){dC(t',t)\over dt}  \ ,
\ee
If at fixed $t'$  we eliminate $t$, we get:
\be
{dR(t',t)\over dC(t',t)}=\beta X(C(t',t)) \  \ \ \ \ {dR\over dC}=\beta X(C) . 
\ee
The plot of $R(t',t)$ versus $ dC(t',t)$ at fixed $t'$ is not a straight line and  becomes
independent of $t'$ for large $t'$.
The function $\beta X(C)$ has the meaning of an effective temperature  at non-equilibrium.

The function $X(C)$ can be computed at equilibrium in a subtle way: it is related both to existence on an
exponential large  number of equilibrium  states and to the overlap distribution at equilibrium.
{If simple aging is valid we have that
\bea
R(t_{w},t) = f(t-t_{w}) \  \mbox{for}\   t-t_{w}>> t_{w}, \\ R(t_{w},t) =
g(t/t_{w})) \  \mbox{for} \ t-t_{w}= O (t_{w}) \  .
\eea
}
In other words, we cool the system at time zero and we make the observations  around a time $t_w$ after cooling:
\begin{itemize}
\item If the scale of time of the observations is small  the results do not depend from the waiting time $t_{w}$, and 
we see the usual microscopical temperature.
\item If the scale of time of the observations is comparable to $t_{w}$, the results depend on the waiting time
and we see an effective temperature that is higher that the microscopic one.
\end{itemize}
{The physical interpretation is simple: the system being at non equilibrium must rearrange itself, and this
produces an extra noise that does not have the adequate correspondence in the response of the system. }

If we are at equilibrium
\be
\lan M(t_{w})M(t)\ran \equiv C(t_{w},t)=C(t-t_{w})
\ee
and only one long experiment is enough.
If not {(we are at off-equilibrium), you have to measure $M$ 100 times over 100 thermal histories to get an error on
$C$ of 10\%.}
In simulations and eventually in optics, you can do measures in parallel: e.g.
\be
{\sum_{i=1,N} \sigma_{i}(t_{w})\sigma_{i}(t) \over N}=C(t_{w},t) \ .
\ee

\subsection{Experimental results}

Using the apparatus described in fig. (\ref{3FLUBIS}), both the fluctuation and the dissipation has been
measured on the same sample \cite{OCIO}. This has been done by cooling a spin glass sample from above to
below the critical temperature. As far as the data show a sizable dependence on the waiting time the authors
of ref. \cite{OCIO} have analyzed  both the raw data have been and the called ageing part, where an
extrapolation to infinite waiting time has been done,

\begin{figure}
\includegraphics[width=0.49\textwidth]{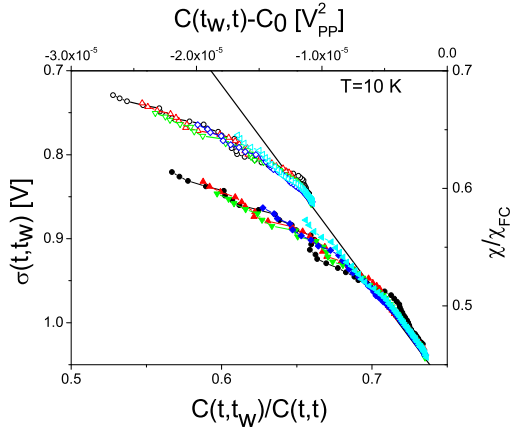}
\caption{
At the left:
Experimental raw results (full symbols) and ageing part (open symbols).
The different curves span the waiting times studied \cite{OCIO}.
}
 \label{AAAAAA}
 \end{figure}

The region where equilibrium FDT does not hold is not horizontal in fig. (\ref{AAAAAA}).
The is not so much curvature in the non-FDT region: theoretical problem: only one temperature, not a continuos
one.  {This is not in agreement with Ising simulation \cite{NUM3}, may be it is in agreement with Heisenberg
simulations.}

{The new off equilibrium fluctuation dissipation relations are extremely important for many reasons:}
\begin{itemize}
  \item They are a first step toward the construction {of a thermodynamics for slightly off equilibrium systems.}
  \item They provide a way to investigated in details {the structure of the different equilibrium states.}
  \item Different theoretical models predict different forms of the function $X(C)$ and in this way {FD
  experiments may better discriminate between different theories.}
 \end{itemize}
 
{It would be extremely interesting to have precise measurements of these off equilibrium fluctuation dissipation relations 
in a variety of different materials as function of the external parameters.}
At the present moment most of the results come from theoretical models or from numerical simulations (in spin
glasses, fragile glasses, colloids and granular materials).

{The effect has been clearly observed in the brilliant experiment in spin glasses for one material.} More
systematic studied are missing. Some effects have been observed in colloids, glasses and granular materials,
but not so clearly as in spin glasses.
{New experimental results are very welcome: they are not easy (the measurement of thermal noise in a
 non-stationary system is a rather difficult enterprise), but the results would be very interesting.}

This difference between the simulations and the experiment may have two different origins:
\begin{itemize}
	 \item The experiment and the numerical simulations do correspond to two different regimes: {the
	 time scales are quite different. } Moreover, if  experimentally we cool an high temperature
	 system, thermalized domains grow with time and the maximum experimental reachable side is about
	 { 100.}
	 In simulations a compact system can be thermalized up to size {20.}
	 \item The numerical simulations are mostly done on Ising systems while the experiment have been
	 mostly done on a more Heisenberg systems with anisotropy an a long range tail of interactions (other
	 systems are more complex); there are indications from other properties that the two systems behave in
	 a different manner.
\end{itemize}

{More extended numerical simulations and  experimental results on other systems are needed
to decide which picture is correct.} One should also consider the possibility that the correlation length in
the equilibrium limit remains fine, but very large (e.g. 1000 lattice units). In such a case one should see
the effects of broken replica symmetries for times of human scale and only for astronomical times the anomaly 
should disappear. The possibility of this phenomenon is difficult to dismiss, but it would not jeopardize the 
interpretation of the experimental data using spontaneously broken replica theory.

{In structural fragile glasses  numerical
simulations and strong theoretical arguments point to the fact that they should belong to case B.}
Unfortunately, although deviations from the equilibrium dissipation fluctuations dissipation relations have
been observed in structural glasses, the situation is not so clear as for spin glasses.

\begin{figure}
\includegraphics[width=0.40\textwidth]{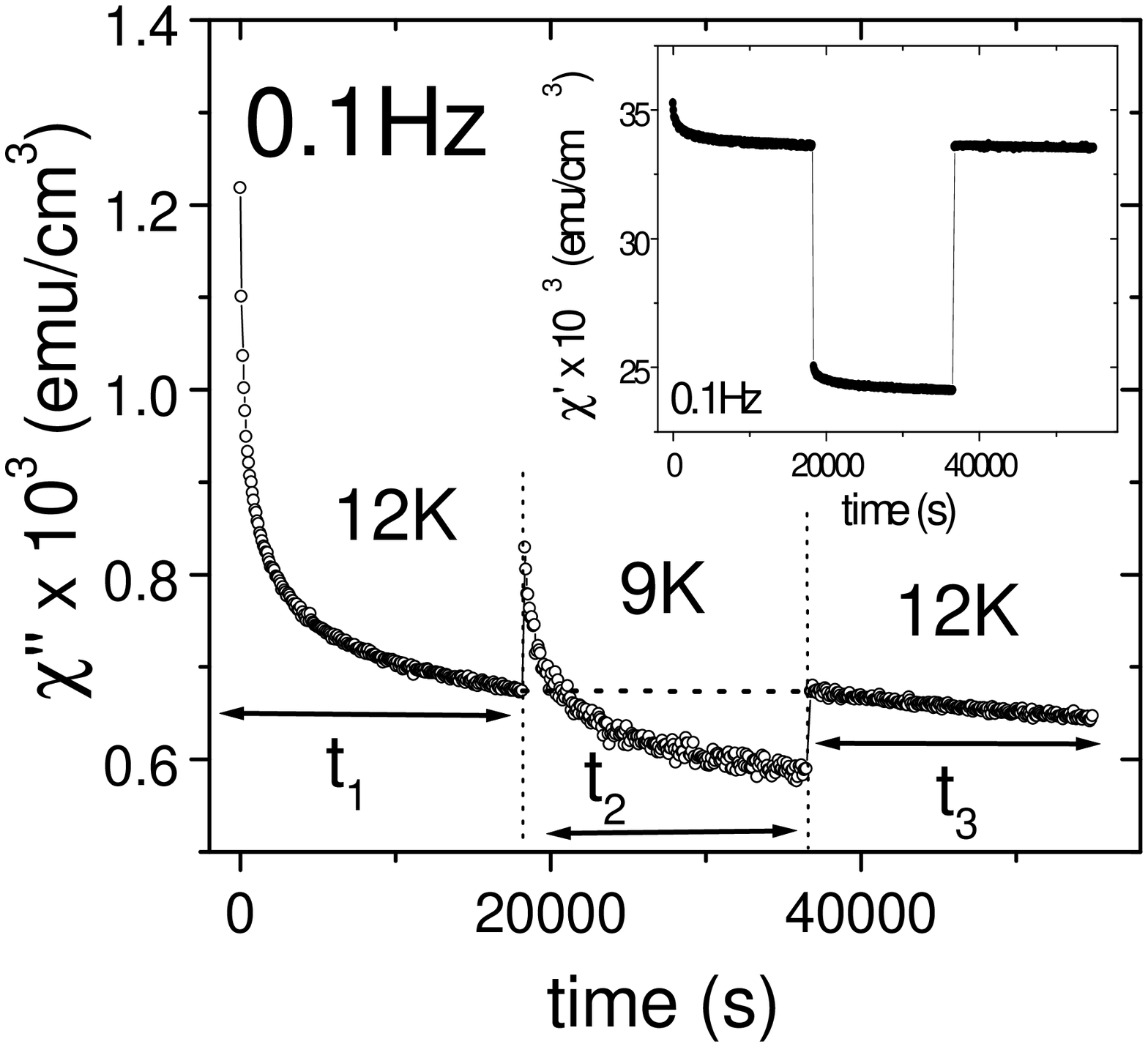} 
\caption{ The system is cooled from an high temperature to 12 K, cooled again to 9~K and heated to 12 K;
{the time spent at 9 K has no effect.} The critical temperature of the sample is 18 K;
$\chi'$ and $\chi''$ are the real and the imaginary part of the susceptibility.
}\label{DD}
\end{figure}

\section{Memory}
\begin{figure}
     \includegraphics[width=0.4\textwidth]{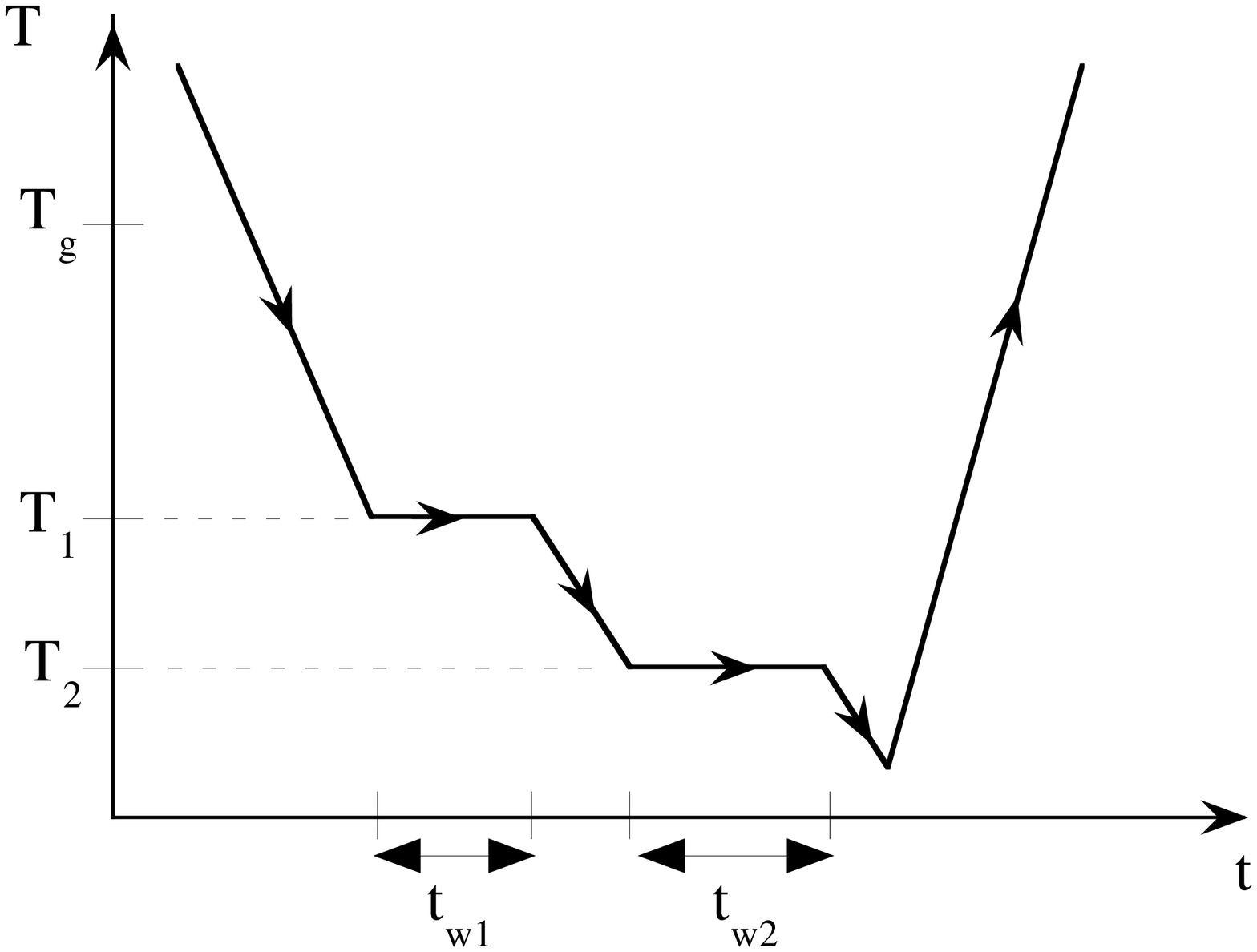} 
 \includegraphics[width=0.4\textwidth]{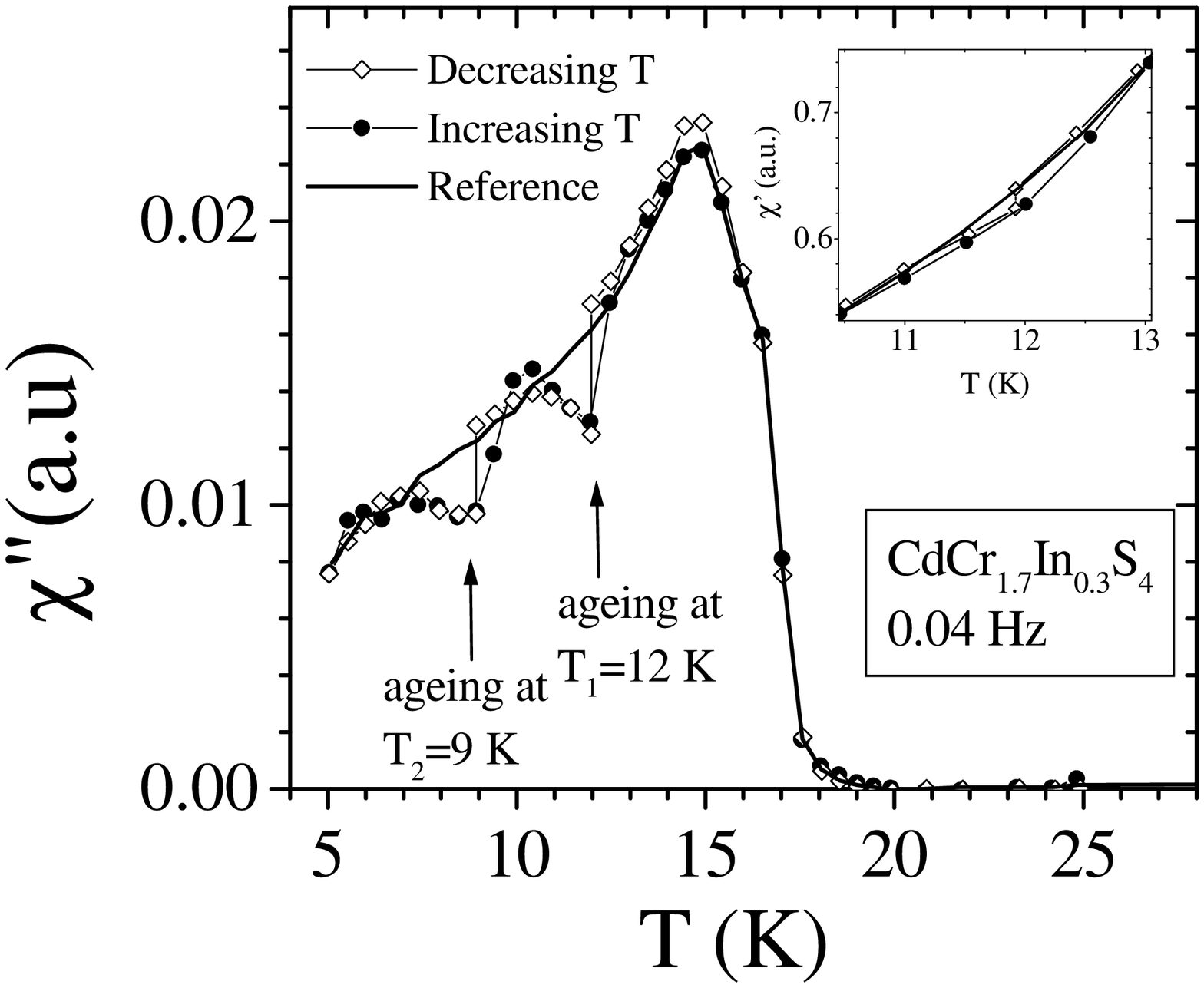} 
 \caption{ In a more complex experiment (temperature dependence in the left panel), you see the dips in $\chi''$
 both on cooling and in heating (right panel).}\label{F}
 \end{figure}

Other impressive phenomena that happen mainly in spin glass are {memory and rejuvenation (or oblivion).}
In the nutshell the magnetic response of the system depends on the history of the system in a very peculiar
way. Some of the relevant experiments are shown in fig. (\ref{DD},\ref{F}).
It is quite remarkable that  we stop at the two temperatures only when cooling, not when heating,
however  one
sees in fig. (\ref{F}) the dips in $\chi''$ during
{\em both}  cooling and in heating.

We cannot discuss in detail these very interesting phenomena. We only remark that 
these effects are more difficult to understand for the following reasons:
\begin{itemize}
    \item {Analytic computations are very difficult} also in mean field theory.
    \item {The effects are much more pronounced for Heisenberg than Ising spins.}
    \item {The effects are nearly invisible in simulations} (mostly done for Ising spins).
 \end{itemize}

A possible explanation could based on two hypothesis:
\begin{itemize}
    \item The free energy landscape at one temperature is not correlated to the free energy
 landscape at an other temperature, {(temperature chaos).}
 \item The degrees of freedom that are {active} at one temperature, are {frozen} at lower temperature
 \end{itemize}

\section{The lowest critical dimension}

{The calculation of free energy increase  due to an interfaces is a
well known method to compute 
the lowest critical dimension for spontaneous symmetry breaking \cite{FPV}.}

In the simplest case we can
consider a system with two possible coexisting phases ($A$ and $B$), with different values of the order
parameter.
For standard ferromagnets
$A$ corresponds to  spins up and $B$ corresponds to  spins down.

{
We will study what happens in a finite system in dimensions $D$ of size
$M^d \times  L$ with  $d=D-1$. } 
We put the system in phase $A$ at
$z=0$ and in phase $B$ at  $z=L$. 
The free energy of the interface is {the
increase in free energy due to this choice of boundary conditions} with
respect to choosing the same phase at $z=0$ and $z=L$ 
In many cases we have that the free energy increase $\delta F(M,L)$
behaves for large $M$ and $L$ as:
{
\be
\delta F(L,M)= M^d/L^\omega \ ,
\ee
where $\omega$ is independent from the dimension. 
}

There is a critical
dimension where the free energy of the interface is finite when both $M$ and $L$ go to infinity at
fixed ratio:
 \be
D_{c}=\omega+1.
\ee
 Heuristic
arguments, which sometimes can be made rigorous,
tell us that when {$D=\omega+1$}, (the lowest critical dimension)
the two phases mix in such a way that symmetry is restored.
{In most cases the value of $\omega$  from mean field theory is the exact
one and therefore we can calculate in this way the value of the
lower  critical dimension. }
The simplest examples are the
ferromagnetic  Ising model $\omega=0$ and the ferromagnetic Heisenberg
model $\omega=1$.

{For spin glasses the order parameter is the overlap $q$.}
We
consider two replicas of the same system described by a Hamiltonian:
\be
H=H[\sigma]+H[\tau] \ ,
\ee
where $H$ is the Hamiltonian of a a single spin glass.
We want to compute the free energy increase corresponding
to imposing an expectation value of {$q$ equal to $q_1$ at
$z=0$ and $q_2$ at $z=L$.}

A complex computation gives (for small $|q_1-q_2|$)
\be
\delta F \propto M L (|q_1-q_2|/L)^{5/2} \propto L^{(D-5/2)} \ .
\ee
{As a consequence, the {\sl naive}
prediction of mean field theory for the lower critical dimension for
spontaneous replica symmetry breaking is $D_c=2.5$.}

These  are testable prediction in Montecarlo on a $L^D$ system.
If the same result is valid also at zero temperature, it may be checked by computing the ground state: One
should find 
\be
\delta E\propto L^{(D-5/2)}\ .
\ee
{I stress that these predictions are {\sl naive}; corrections to the mean
field theory are neglected, however n known cases these kind of computations give the correct result.}

\section{Conclusions}
A strong effort should be done in order to have better quantitative predictions and a more
precise comparison between theory and experiments. 
There are open problems in spin glasses, glasses and related fields where I would like to see
a progress:
\begin{itemize}
	 \item {A systematic and careful experimental study of the fluctuation-dissipation relations is
	 needed.}
	 \item { It would be important to obtain more precise understanding of the glass transition in
	 structural fragile glass, and to get quantitative predictions on the behavior of the dynamics in the
	 region where it is dominated by escapes from barriers.}
	 \item {New results should be obtained of the  form of time behavior in {\it non-equilibrium
	 state}, where only partial results are known.}
	 \item {In order to arrive to a precise assessment of the behavior of three dimensional spin
	 glasses it would be very important to develop further the renormalization group approach and/or
	  other theoretical or experimental techniques.}
\end{itemize}

\end{document}